# Metal hydrides achieve high-$T_c$ superconductivity at low pressure by mimicking high-pressure H₃S chemical bonding

Wendi Zhao[1], Shumin Guo[1], Chengda Li[1], Abhiyan Pandit[3], Tian Cui[2,1#], Defang Duan[1]*, and Maosheng Miao[3,4]†

[1]State Key Laboratory of High Pressure and Superhard Materials, Key Laboratory of Material Simulation Methods & Software of Ministry of Education, College of Physics, Jilin University, Changchun, 130012, China

[2]Institute of High Pressure Physics, School of Physical Science and Technology, Ningbo University, Ningbo 315211, China

[3]Department of Chemistry and Biochemistry, California State University, Northridge, California 91220, USA

[4]Department of Earth Science, University of California Santa Barbara, California 93106, USA

Corresponding authors.
*duandf@jlu.edu.cn
 # cuitian@nbu.edu.cn
 †mmiao@csun.edu

## Abstract

Compressed hydrides are promising candidates for high-temperature superconductivity, yet achieving simultaneous structural stability and high-$T_c$ at low pressures remains challenging. Here, we introduce a new mechanism for accomplishing this goal by mimicking the bonding characteristics of high-pressure $H_3S$ within metal hydrides. Using $Li_3CuH_4$ as an example, its Cu-H covalent interaction effectively mimics the core function of the S-H bonding in $H_3S$. This interaction not only induces a high hydrogen-derived electronic density of states at the Fermi level, but also softens the hydrogen phonon modes, thereby significantly enhancing the electron-phonon coupling. Furthermore, embedding the strongly ionic $Li_3H$ lattice into the covalent Cu-H framework stabilizes the structure at significantly low pressures via a chemical-template effect, while maintaining high-$T_c$. $Li_3CuH_4$ exhibits excellent thermodynamic stability at 20 GPa, with a $T_c$ of 39.25 K at 12 GPa. Further comprehensive high-throughput studies on $Li_3MH_4$ (M = transition metal) compounds uncover general principles applicable to a broader range of compounds. This work establishes a new paradigm for the simultaneous optimization of the stability and high-temperature superconductivity of metal hydrides through complementary sublattice interactions, thus advancing the search for practical and viable superconducting materials.

**Introduction**

Compressed hydrides have recently garnered extensive attention as a promising class of high-temperature superconductors, leading to substantial research progress and groundbreaking records in superconducting transition temperatures[1, 2]. Cubic $H_3S$ was the first hydride superconductor experimentally confirmed to exhibit remarkable superconductivity, with a record-high $T_c$ of 203 K at 155 GPa[3-5]. Subsequently, the clathrate hydride $LaH_{10}$ achieved a $T_c$ of up to 260 K at 170 GPa[6-11]. These discoveries have significantly propelled the exploration of superconducting hydrides, expanding the scope to ternary hydride systems. Theoretical predictions suggest that some ternary hydrides, such as $Li_2NaH_{17}$[12] and $LaSc_2H_{24}$[13], exhibit superconducting properties that potentially exceed room temperature. However, these high-$T_c$ hydrides are typically stable only above 150 GPa, which poses significant challenges for experimental synthesis and hinders the production of large-sized samples for practical applications.

Recent research has intensified efforts to lower the stabilization pressures of superconducting hydrides[14-20]. A widely adopted strategy involves using prototypical high-$T_c$ hydrides such as $LaH_{10}$ and $H_3S$ as parent structures and incorporating additional elements to modulate chemical bonding, thereby enhancing stability. For instance, $LaBeH_8$ can be viewed as a derivative of $LaH_{10}$[14, 15]. Significantly different from its parent, $LaBeH_8$ features a fluorite-type alloy framework formed by Be and H atoms, which stabilizes at substantially lower pressures than the pure hydrogen framework in $LaH_{10}$. Theoretical predictions indicate that $LaBeH_8$ becomes thermodynamically stable above 98 GPa, and the measured $T_c$ at 80 GPa is 110 K[14, 15]. Similarly, isostructural compounds such as $ThBeH_8$, $CeBeH_8$, and $BaSiH_8$ also stabilize at moderate pressures and exhibit promising superconducting properties[16-18]. These results demonstrate that the stabilization pressure of parent hydrides can be substantially reduced in ionic clathrate hydrides through rational elemental substitution. However, these similar strategies do not work well for the $H_3S$ system. Researchers have extensively explored doping or elemental substitution in $H_3S$, typically with additional nonmetallic elements, to enhance stability or superconductivity. Nevertheless, most predicted derivatives of $H_3S$ still require high pressures for stability. Examples include $H_3Se$ ($T_c$ = 110 K at 200 GPa) [21], $H_6SSe$ ($T_c$ = 196 K at 200 GPa) [22], and $H_6SP$ ($T_c$ = 159 K at 200 GPa) [23]. Crucially, thermodynamic stability–indicating that a structure resides at a global energy minimum and resists transformation or decomposition–is essential for experimental synthesis. Although some $H_3S$ derivatives remain dynamically stable at low pressures, their thermodynamic stability requires high pressures, or they exist in a metastable state. For instance, $H_6SCl$ and $H_6SBr$ are dynamically stable down to around 5 GPa, yet their thermodynamic stability[24, 25] still requires pressures above 200 GPa. These results indicate the inherent difficulty in achieving both low-pressure stability and superconductivity in nonmetallic hydrides possessing $H_3S$-like structures.

In light of these limitations, we propose the exploration and design of metal hydrides that mimic the covalent framework of $H_3S$, aiming to achieve both thermodynamic stability and superconductivity at significantly reduced pressures. Interestingly, the cubic perovskite structure exhibits inherent similarities to the sublattice of $H_3S$. Some perovskite hydrides such as $KInH_3$[26] and $KGaH_3$[27] are predicted to exhibit superconducting potential at moderate pressures, although they remain thermodynamically metastable under these conditions. This suggests that the pursuit of metal hydrides with $H_3S$-like frameworks requires designing novel structural motifs. Our approach represents a new and largely unexplored direction, providing a promising alternative to previous efforts focused on tuning the modifying structure of $H_3S$ itself.

In this work, we propose novel strategies to achieve simultaneous structural stability and high-temperature superconductivity in metal hydrides at low pressures by mimicking key bonding characteristics of high-pressure $H_3S$. Taking $Li_3CuH_4$ as a representative example, this compound is composed of two sublattices, $CuH_3$ and $Li_3H$, with complementary chemical interactions. The Cu-H bonds display covalent interactions at moderate pressure, analogous to the S-H bonds in $H_3S$ at much higher pressures. This effectively promotes high hydrogen-derived electronic density of states (DOS) at the Fermi level and significantly enhances the electron-phonon coupling (EPC). Concurrently, the covalent Cu-H framework can be stabilized through intercalation with a strongly ionic $Li_3H$ lattice via a chemical-template effect. The additional charge donated by Li atoms further increases the electron density on hydrogen, thereby enhancing superconducting properties. Based on this complementary lattice interaction, $Li_3CuH_4$ simultaneously maintains excellent thermodynamic stability and high-temperature superconductivity at low pressure. Our findings demonstrate that rationally combining metal elements with suitable atomic radii and electronegativities effectively balances ionic and covalent interactions within the host lattice, enabling synergistic optimization of structural stability and superconductivity at low pressures.

**Results and discussion**

The key to mimicking the superconductivity of $H_3S$ lies in designing a compound that shares similar structural and electronic properties with its superconducting phase. At pressures exceeding 180 GPa, $H_3S$ crystallizes in a cubic structure with *Im*-3*m* symmetry[3]. It can be viewed as consisting of two interpenetrating [$SH_3$] sublattices (Figure 1a), where the covalent bonds between sulfur and hydrogen atoms not only stabilize the crystal structure but also induce strong electron-phonon coupling, which is crucial for achieving high-temperature superconductivity. Therefore, an ideally substituted material must establish analogous robust bonding interactions. We note that, under high pressures corresponding to superconductivity, $H_3S$ undergoes charge reversal; sulfur atoms become positively charged (oxidized), and hydrogen atoms become negatively charged (reduced) (Figure 1b). To replicate superconducting $H_3S$, one must select a compound in which hydrogen naturally acquires a negative charge at lower pressures. Metal-hydrogen compounds inherently satisfy this criterion. Particularly, late transition metals emerge as ideal candidates, as they not only donate electrons to hydrogen but also possess electronegativity values close to that of hydrogen.

This similarity facilitates orbital hybridization and enables the formation of covalent bonding interactions, thereby effectively simulating the bonding behavior of S–H under high pressure.

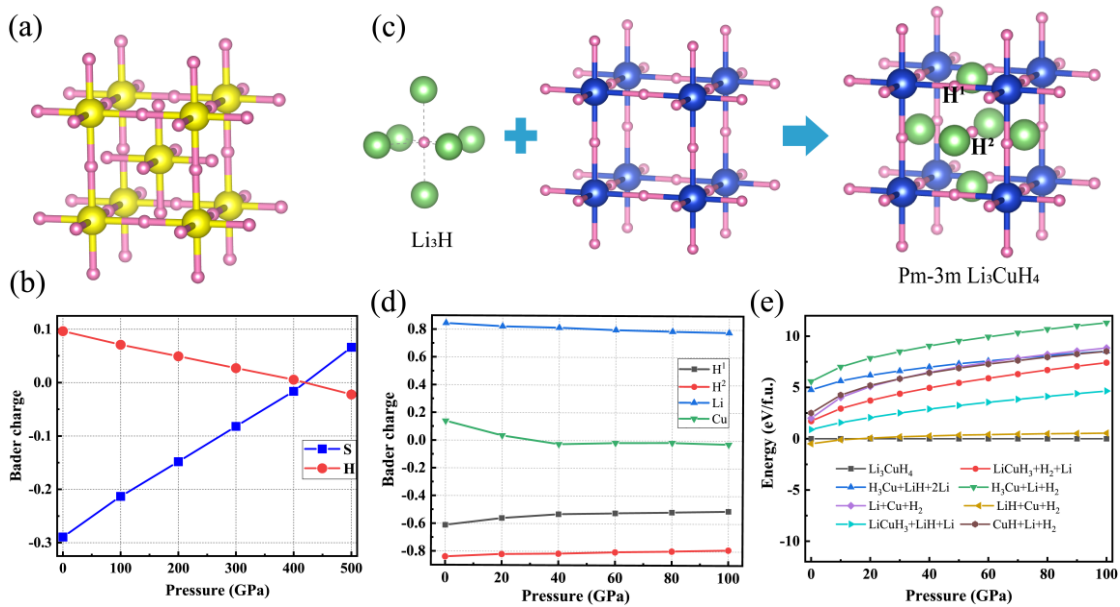

**Figure 1** (a) Crystal structures of $Im$-$3m$ $H_3S$. (b) Calculated Bader charges of H and other atoms in $SH_3$. (c) Structural evolution: A cubic sublattice of $H_3S$ combined with a $Li_3H$ sublattice forms $Pm$-$3m$ $Li_3CuH_4$. (d) Calculated Bader charges of H and other atoms in $Li_3CuH_4$. (e) Calculated formation enthalpy of $Li_3CuH_4$ relative to various decomposition pathways as a function of pressure.

Notably, simultaneously replicating both sublattices in $H_3S$ poses significant challenges. Although directly replacing sulfur in $H_3S$ with transition metals is conceptually feasible—for instance, forming $H_3Cu$—such structures are generally energetically unfavorable and difficult to stabilize (see Figure S1). Instead, a more effective strategy involves replicating one essential covalent sublattice and integrating it with an ionic sublattice formed by other highly electropositive elements (e.g., alkali/alkaline earth metals) to achieve complementary chemical interactions. For example, in the constructed $Li_3CuH_4$, hydrogen atoms occupy distinct Wyckoff positions: the $H^1$ atoms at the 3d site (0, 0.5, 0) coordinates octahedrally with Cu atoms, forming the covalent $CuH_3$ sublattice. The $H^2$ atoms at the 1b site (0.5, 0.5, 0.5) resides at the center of a Li octahedron, constituting the ionic $Li_3H$ sublattice (Figure 1c). Bader charge analysis reveals that Li donates more electrons to hydrogen, with the $H^2$ atoms acquiring a significantly higher electron count compared to $H^1$. Notably, as pressure increases, the electron donation from Cu to hydrogen decreases, indicating more pronounced covalent interactions with neighboring $H^1$ atoms (Figure 1d). Consequently, in $Li_3CuH_4$, the $Li_3H$ sublattice leverages the strong electropositivity of lithium to donate sufficient electrons to hydrogen, not only enhancing electrostatic interactions but also serving as a "chemical template" that supports the structural stability of the $CuH_3$ framework. Thermodynamic analysis reveals that $Li_3CuH_4$, formed through this combined sublattice approach, is energetically more favorable than $LiCuH_3$, which results from introducing a single Li atom into $CuH_3$. More specifically, we further decomposed the formation enthalpy (H) contributions into internal energy (U) and PV work: $H = U + PV$. $Li_3CuH_4$ exhibits a significantly lower $\Delta U$ compared to the reference state ($LiCuH_3$ + LiH + Li), and its $\Delta PV$ term also decreases progressively with increasing pressure. Analysis of enthalpy curves (Figure 1e), coupled with extensive random structure searches involving over 20,000 distinct configurations within the ternary Li-Cu-H phase diagram, demonstrates that $Li_3CuH_4$ maintains thermodynamic stability at 20 GPa (see Figure S2).

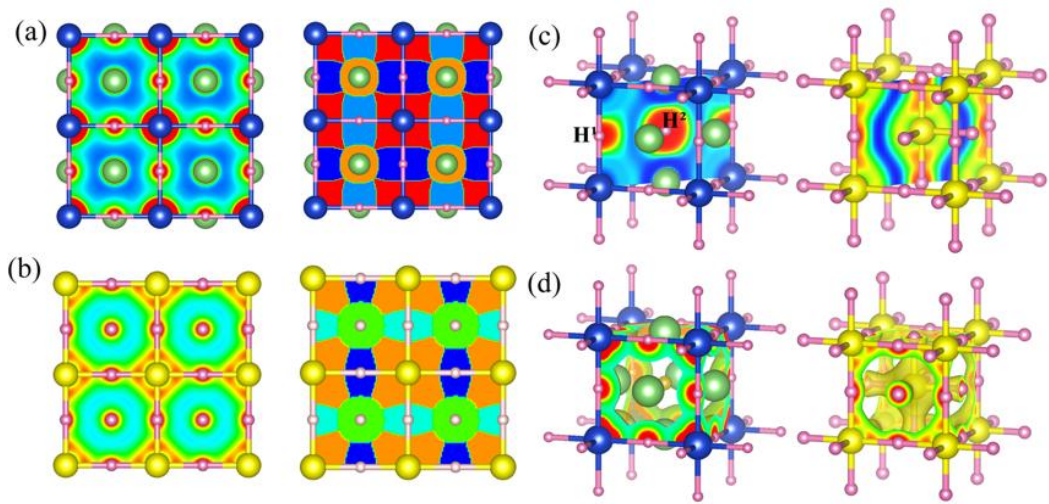

**Figure 2** Charge density (left) and Bader charge volume (right) of (a) $Li_3CuH_4$ and (b) $H_3S$. (c) Electron localization function (ELF) and (d) Partial charge density within the energy window [$E_f$-1, $E_f$] eV of $Li_3CuH_4$ (left) and $H_3S$ (right). The calculations were performed using the structures of $Li_3CuH_4$ and $H_3S$ at 12 GPa and 200 GPa, respectively.

Subsequently, we further elucidate the similarities in chemical bonding between $Li_3CuH_4$ and $H_3S$. The significant charge density distribution along the Cu-H bond in $Li_3CuH_4$ indicates the presence of orbital hybridization interaction. We further employed Bader volume analysis to partition the electron density (Figure 2a-b). This method defines atomic volumes by identifying local maxima in the electron density (typically corresponding to atomic nuclei), with boundaries delineated by zero-flux surfaces in the electron density gradient. Cu has a large Bader volume due to its extended 3d orbital and hybridization with the H 1s orbital. This charge density distribution pattern exhibits a striking similarity to the covalent S-H bonds found in $H_3S$. In contrast, typical ionic crystals such as LiH feature $H^-$ ions that have acquired substantial electron density, and thus exhibit a large Bader volume (see Figure S3). Electron localization function (ELF) analysis shows weak electron localization between Cu and H atoms (Figure 2c), supporting the presence of covalent bonding. In contrast, almost no electron localization is observed between Li and H atoms, confirming Li-H ionic bonding. Differential charge density analysis clearly highlights regions of higher charge accumulation: the $H^2$ site exhibits significantly more charge enrichment than the $H^1$ site, and charge accumulation is also evident near the Cu atoms (see Figure S4). Collectively, these multifaceted results demonstrate the coexistence of covalent Cu-H bonding and strongly ionic Li-H bonding within $Li_3CuH_4$. This complementary bonding interaction synergistically enhances the stability of the structure. While the covalent character of the Cu-H interaction is weaker than that of the S-H bond in $H_3S$, it nonetheless exerts a significant influence on the lattice charge distribution in the metal hydride. The partial charge density distribution near the Fermi level originates primarily from the covalent Cu-H bonds, particularly the enhanced hydrogen-derived charge density (Figure 2d). This plays a key role in strengthening electron-phonon coupling (EPC), functionally underscoring the core similarity to the S–H covalent bonds in $H_3S$.

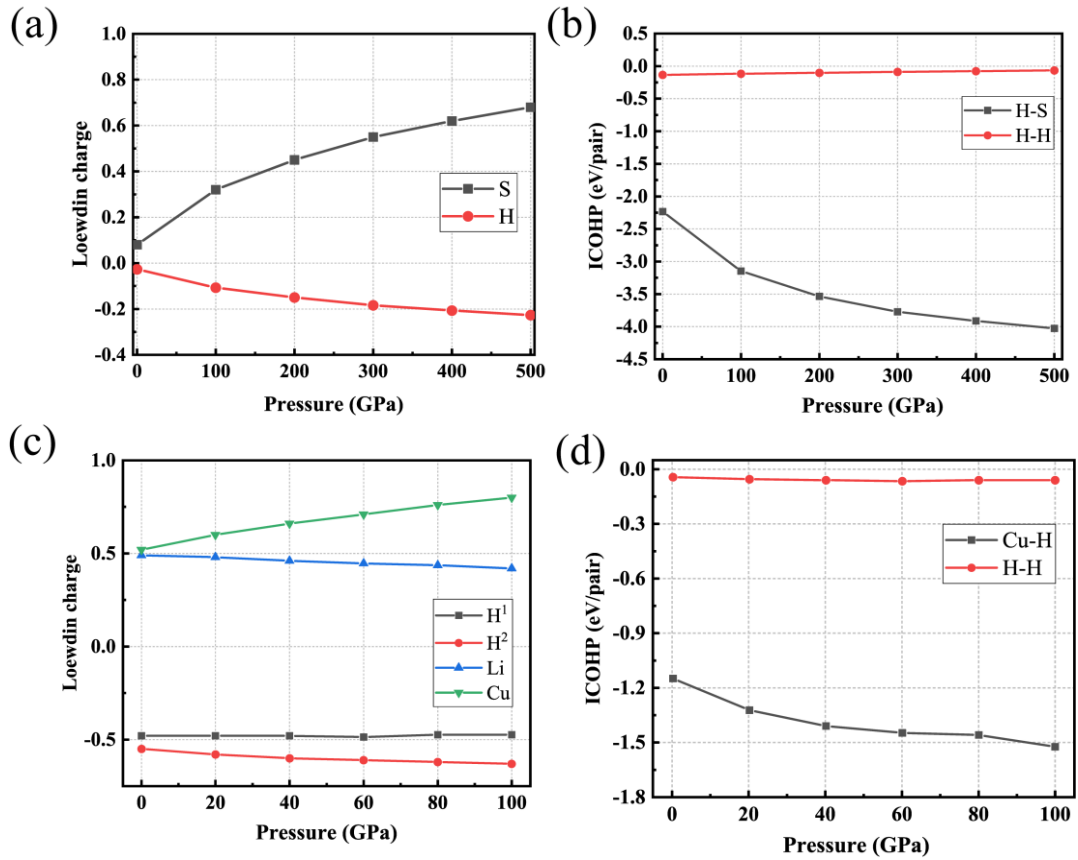

**Figure 3** Pressure-dependent Loewdin charge and Integrated Crystal Orbital Hamiltonian Population (ICOHP) for (a-b) $H_3S$ and (c-d) $Li_3CuH_4$.

We investigated the pressure dependence of Loewdin populations[28] and Integrated Crystal Orbital Hamiltonian Population (ICOHP) values[29] in $H_3S$ and $Li_3CuH_4$ (Figure 3). The Loewdin charge distribution is closely related to orbital hybridization and covalent bond strength. Increased pressure shortens S-H bond lengths and enhances covalent interactions, thereby elevating hydrogen population contribution in hybrid orbitals and consequently increasing H Loewdin charge. The strengthening of S-H covalent bonding under pressure is further reflected in the progressively more negative ICOHP (values. Similarly, in $Li_3CuH_4$, increasing Loewdin populations on Cu and the corresponding rise in |ICOHP| values with pressure demonstrate the strengthening of Cu-H covalent bonds. Conversely, robust ionic bonding persists between Li and H atoms, as hydrogen readily acquires sufficient electrons from lithium. These results demonstrate that $Li_3CuH_4$ effectively replicates key electronic properties of high-pressure $H_3S$ at substantially lower pressures, holding promise for inducing strong electron-phonon coupling.

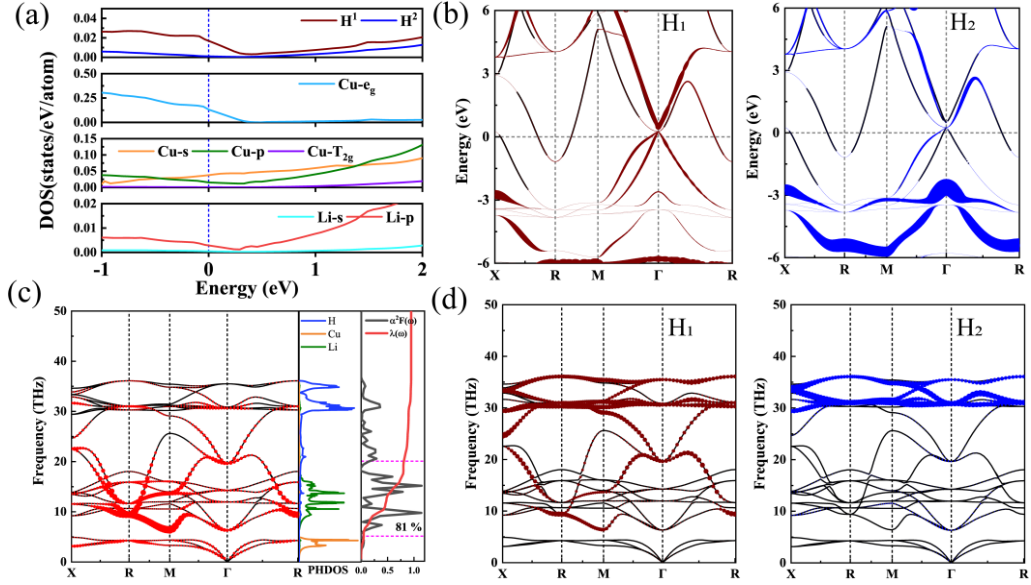

**Figure 4** (a) Each atom-projected density of states (PDOS) and (b) Band structure with specific H-site ($H^1/H^2$) weights for $Li_3CuH_4$ at 12 GPa. (c) Phonon dispersion curves with projected phonon density of states (PHDOS), Eliashberg spectral function $\alpha^2 F(\omega)$, and electron-phonon coupling integral $\lambda(\omega)$. (d) Site-resolved phonon dispersion projections for $H^1/H^2$ atoms. All data calculated at 12 GPa.

Figures 4a-b present the electronic band structure and density of states (DOS) of $Li_3CuH_4$. A key feature is the significantly higher density of states (DOS) at the Fermi level per $H^1$ atom compared to $H^2$ atoms, primarily attributed to orbital hybridization between the Cu d orbitals and the $H^1$ 1s orbitals. Concurrently, the extended Cu 4s states exhibit considerable overlap with $H^1$ 1s orbitals, and the strong Pauli repulsion between them pushes $H^1$ electrons into higher-energy electronic states. In contrast, the predominantly ionic interaction between Li and $H^2$ atoms results in $H^2$ ions occupying lower-energy valence band states, leading to negligible DOS contributions at the Fermi level. Figure 4c presents the phonon dispersion curves with projected phonon density of states (PHDOS), Eliashberg spectral function $\alpha^2 F(\omega)$, and electron-phonon coupling integral $\lambda(\omega)$ of $Li_3CuH_4$. Phonon vibrational modes of Cu and Li atoms are distributed in low- and mid-frequency regions, respectively, while light H atoms exhibit phonon modes associated with mid- and high-frequency vibrations. Notably, the softened mid-frequency phonon modes involving H atoms significantly enhance electron-phonon coupling (EPC), as evidenced by pronounced peaks in α²F(ω) and marked increase in λ(ω). The relative contributions of phonon modes to EPC are visualized by red dots scaled according to their coupling strength. Dense red dots clustered in the mid-frequency softened phonon modes highlight their dominant role, collectively contributing 81% to the total EPC. Notably, $H^1$ and $H^2$ atoms exhibit distinct phonon vibrational characteristics: the softened optical modes primarily originate from $H^1$ atoms (Figure 4d), underscoring their predominant contribution to EPC. The EPC parameter λ can be expressed as[30]:

$$\lambda = \frac{\eta}{M\langle\omega^2\rangle} = \frac{N(\varepsilon_F)\langle I^2\rangle}{M\langle\omega^2\rangle}$$

where $N(\varepsilon_F)$, $\langle I^2\rangle$, and $\langle\omega^2\rangle$ represent the density of states at the Fermi level, averaged electron-phonon matrix elements, and averaged phonon frequencies, respectively. $H^1$ atoms exhibit substantially higher DOS at the Fermi level compared to $H^2$, while Cu-$H^1$ orbital hybridization reduces the phonon frequencies of $H^1$. Notably, the frequency range of the soft phonon mode associated with $H^1$ is very similar to the phonon frequency driving the $H_3S$ strong electron phonon coupling[3-5]. Simultaneously, ionic interactions between Li and H atoms induce localized high-frequency vibrational modes, characteristic of ionic crystals. Although these high-frequency modes contribute weakly to electron-phonon coupling (EPC), the strong electrostatic interactions significantly enhance structural stability. $Li_3CuH_4$ exhibits a $T_c$ of up to 39.25 K at a moderate pressure of 12 GPa. Therefore, the complementary bonding interactions between Li-H and Cu-H constitute the key mechanism enabling $Li_3CuH_4$ to replicate both the electronic properties and phonon characteristics of high-pressure $H_3S$, thereby achieving remarkable superconductivity under low-pressures.

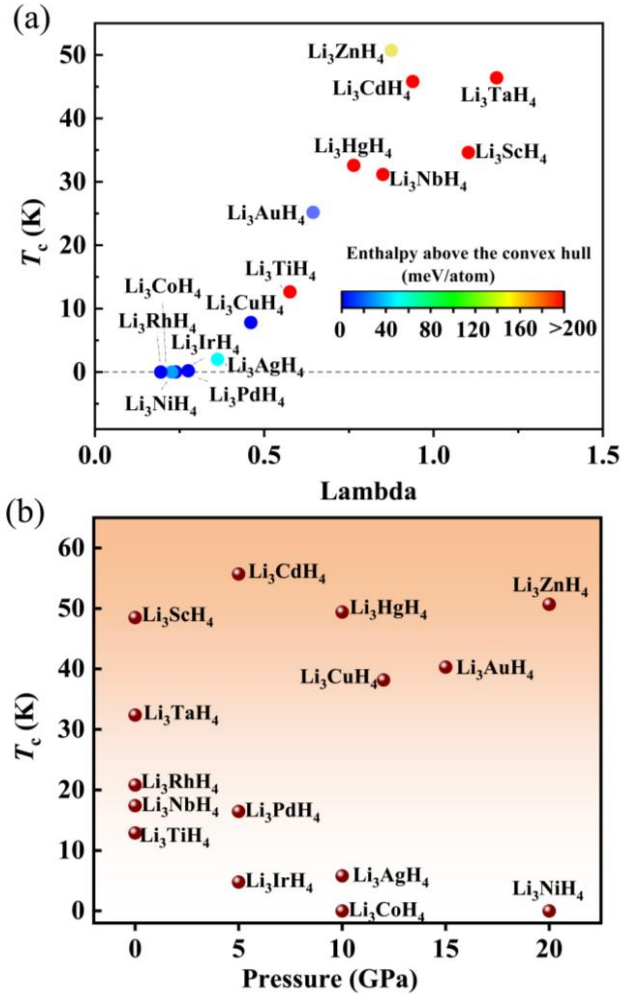

**Figure 5** (a) High-throughput computational screening of $Li_3MH_4$ (M = transition metal) compounds at 20 GPa. The colored dots represent the thermodynamic stability (distance from the convex hull) at 20 GPa. (b) The $T_c$ values of some $Li_3MH_4$ systems at dynamically stable critical pressures.

Building upon the $Li_3CuH_4$ structural prototype, we systematically explored the $Li_3MH_4$ (M = transition metal) system via high-throughput computations to evaluate compositional effects on stability and superconductivity, uncovering universally applicable principles for hydride design. Candidate structures were rigorously evaluated for thermodynamic and dynamic stability at 20 GPa, identifying 15 dynamically stable configurations (Figure 5). The valence electron count, electronegativity, and atomic radius of the transition metals are critical factors governing both the structural stability and superconductivity in $Li_3MH_4$ systems (Figure S6). Significantly, late transition metals possess electronegativity close to that hydrogen, enabling an effective balance of the covalent-ionic bond interactions within the host lattice. Consequently, their corresponding structures demonstrate superior thermodynamic stability—either lying on the convex hull or located within 80 meV/atom above it. Four thermodynamically stable phases (M = Cu, Rh, Pd, Ni) were confirmed at 20 GPa, with their dynamical stability onset pressures all below this value (see Figure S7). $Li_3RhH_4$ and $Li_3PdH_4$ exhibit $T_c$ values of 21 K and 16 K at ambient pressure and 5 GPa, respectively. Notably, several metastable phases show enhanced superconducting performance. $Li_3ScH_4$ and $Li_3TaH_4$ achieve high $T_c$ values of 49 K and 32 K at ambient pressure, while $Li_3ZnH_4$ and $Li_3CdH_4$ reach 51 K and 56 K at 20 GPa and 5 GPa, respectively. These metastable systems feature relatively longer M-H bond lengths compared to stable phases, which can be attributed to lattice expansion caused by the larger atomic radii of the M elements. This structural softening effect enhances the phonon contribution to electron-phonon coupling (EPC).

It is worth emphasizing that $Li_3CuH_4$ served as a paradigm to elucidate the mechanism for simulating the structural and electronic properties of cubic $H_3S$ at low pressures. The core objective of this mimicry is to capture the essential features of S-H covalent bonding in $H_3S$ that drive high-temperature superconductivity, rather than being limited to absolute structural replication. Our extensive structural search within the Li–Cu–H system uncovered additional thermodynamically stable phases (e.g., $LiCu_3H_4$, $LiCuH_2$, $LiCu_3H_3$), all of which feature Cu–H covalent frameworks that emulate S–H bonding characteristics while exhibiting robust stability and superconductivity (see Figures S2, S8–S9). Specifically, $LiCu_3H_3$ remains thermodynamically stable at 15 GPa, and $LiCuH_2$ at 30 GPa, with predicted $T_c$ values reaching 35 K at 6 GPa and 46 K at 7 GPa—pressures corresponding to their critical dynamically stable points, respectively. Given the diversity in stoichiometry and elemental chemistry across hydride systems, structures incorporating bonding motifs that effectively simulate the covalent character of S-H bonds may exist in various configurations. This represents a promising new direction that warrants further exploration.

**Conclusions**

In summary, our work establishes a new mechanism for synchronizing structural stabilization and superconductivity enhancement in metal hydrides at low pressures, centered on effectively replicating the essential bonding characteristics of high-pressure $H_3S$ through complementary sublattice interactions. In $Li_3CuH_4$, the covalent Cu-H interaction specifically replicates the functional role of the S-H bond in $H_3S$. This not only promotes the formation of a DOS with high hydrogen-composition at the Fermi level but also softens hydrogen-related phonon modes, significantly enhancing EPC. Crucially, embedding strongly ionic $Li_3H$ lattice within the Cu-H framework facilitates complementary synergy between covalent and ionic sublattices via a chemical templating effect, enabling high-$T_c$ superconductivity while maintaining structural integrity. $Li_3CuH_4$ is thermodynamically stable at a moderate pressure of 20 GPa and achieves a $T_c$ up to 39.25 K at 12 GPa. High-throughput screening further reveals that late transition metals are ideal candidates, as they facilitate electron transfer to hydrogen at modest pressures while forming stable covalent bonds—behavior closely analogous to S–H bonding in high-pressure $H_3S$. This mechanism provides an unprecedented pathway to simultaneously achieve stability and high-$T_c$ superconductivity in a wide range of hydrides with metal-H covalent bonds, thereby establishing a new paradigm for the search of practical high-temperature superconductors and advancing the design of unconventional metal alloy compounds that exhibit exceptional properties grounded in predictable bonding principles.

**Declaration of competing interest**

The authors declare that they have no known competing financial interests or personal relationships that could have appeared to influence the work reported in this paper.

**Acknowledgements**

This work was supported by the National Key Research and Development Program of China (No. 2022YFA1402304 and 2023YFA1406200), the National Natural Science Foundation of China (Grant Nos. 12122405 and 12274169), the Program for Science and Technology Innovation Team in Zhejiang (No. 2021R01004), and the Fundamental Research Funds for the Central Universities. A.P. and M.M. acknowledge the support of the DoD Research and Education Program for Historically Black Colleges and Universities and Minority-Serving Institutions (HBCU/MI) Basic Research Funding under grant No. W911NF2310232. M.M. also acknowledges the support of the National Science Foundation (NSF) under grant nos. DMR-1848141 and OAC-2117956, the Camille and Henry Dreyfus Foundation, and the California State University Research, Scholarship, and Creative Activity (RSCA) award. Some of the calculations were performed at the High-Performance Computing Center of Jilin University, using TianHe-1(A) at the National Supercomputer Center in Tianjin.

060102.

22. Liu B, Cui W, Shi J *et al.* Effect of covalent bonding on the superconducting critical temperature of the H-S-Se system. *Phys Rev B*. 2018; **98**(17): 174101.
23. Shao Z, Song H, Yu H *et al.* Ab Initio Investigation on the Doped $H_3S$ by V, VI, and VII Group Elements Under High Pressure. *J Supercond Nov Magn*. 2022; **35**(4): 979-986.
24. Hai Y-L, Tian H-L, Jiang M-J *et al.* Prediction of high Tc superconductivity in $H_6SX$(X=Cl,Br) at pressures below one megabar. *Phys Rev B*. 2022; **105**(18): L180508.
25. Huo Z, Duan D, Jiang Q *et al.* Cubic $H_3S$ stabilized by halogens: High-temperature superconductors at mild pressure. *Sci China Phys Mech Astron*. 2023; **66**(11): 118211.
26. Cerqueira TF, Fang YW, Errea I *et al.* Searching Materials Space for Hydride Superconductors at Ambient Pressure. *Adv Funct Mater*. 2024: 2404043.
27. Du M, Huang H, Zhang Z *et al.* High-Temperature Superconductivity in Perovskite Hydride Below 10 GPa. *Adv Sci*. 2024; **11**(42): e2408370.
28. Müller PC, Ertural C, Hempelmann J *et al.* Crystal Orbital Bond Index: Covalent Bond Orders in Solids. *J Phys Chem C*. 2021; **125**(14): 7959-7970.
29. Nelson R, Ertural C, George J *et al.* LOBSTER: Local orbital projections, atomic charges, and chemical-bonding analysis from projector-augmented-wave-based density-functional theory. *J Comput Chem*. 2020; **41**(21): 1931-1940.
30. Allen PB, Dynes RC. Transition temperature of strong-coupled superconductors reanalyzed. *Phys Rev B*. 1975; **12**(3): 905-922.